\documentstyle{mn}

\newif\ifAMStwofonts

\input psfig.h
\input macros.h
\input mn.h
\input psfig.h
\title{Discordant redshifts in compact groups}
\author[A. Iovino and P. Hickson]
       {A.~Iovino,$^1$
	and P.~Hickson,$^{1,2}$\\
        $^1$Osservatorio Astronomico di Brera, Via Brera 28, I-20121 Milano, 
	Italy \\
        $^2$Department of Geophysics and Astronomy, University of British 
	Columbia, 2219 Main Mall, Vancouver, BC V6T 1Z4, Canada
	}

\begin{document}

\maketitle

\begin{abstract}
This paper addresses the long standing 
question of discordant redshifts in compact groups. 
We have used an homogenous catalogue of 173 compact groups selected 
by an automated procedure to objectively predict the fraction of
discordant redshifts with high statistical 
accuracy, and then applied these results to the sample 
of 92 compact groups in Hickson's revised catalogue. 
Our results confirm that projection effects alone can account for the high
incidence of discordant redshifts in compact groups. 
We have also examined the spatial distribution of discordant 
galaxies in Hickson's compact groups. Contrary to
previous studies, we find that there is no evidence for central
concentration of discordant galaxies.

\end{abstract}


\section{Introduction}

Compact groups of galaxies pose a number of interesting questions for
astronomers. The most long-standing and controversial is
that of discordant redshifts. The difficulty began when the first
two known compact groups, Stephan's Quintet (Stephan 1877) and
Seyfert's Sextet (Seyfert 1948a,b), were both found to contain a
galaxy whose redshift differed greatly from that of the other
group members (Burbidge \& Burbidge 1961). This surprising result
was repeated with the discovery (Sargent 1968) of a discordant redshift
in VV172 (Vorontsov-Vel'yaminov 1959), and in many more compact groups
(Hickson \etal\ 1992).

Because of the seemingly high incidence of discordant redshifts,
doubts have been expressed as to whether they can be
explained by projection effects alone (Arp 1973, Sulentic 1984). 
Early attempts to answer this question have not been conclusive. From a
study of $\sim 200$ galaxy quartets and triplets, Rose (1977) concluded that 
projection effects were responsible for the discrepant redshifts 
cases observed. However, his conclusion was
based on a sample of galaxies with few measured velocities. Sulentic
(1987) reached the opposite conclusion, based on counts of galaxies
around Hickson's (1982) sample of compact groups. However this work
underestimated the probability of background contamination by
requiring the discordant galaxy to be inside the group and not just 
outside (still making it an isolated and compact group). Hickson \etal\
(1988) computed separately probabilities for internal and external 
discordant redshifts. They concluded that the overall number of
discordant redshifts was consistent with projection effects, but that
there was (at the 98\% level) an excess of internal discordant redshifts.
Mendes de Oliveira (1995) reached a similar conclusion, using more
complete redshift data, and suggested that the apparent 
concentration of discordant redshifts towards the centres of the
groups might be due to weak gravitational lensing (see also Mendes de Oliveira \&
Giraud 1994).

These recent studies compare the number of discordant quintets with the 
number expected by chance, based on the number of accordant quartets in
the catalogue and the surface density of field galaxies. Since the 
number of groups studied is small, the statistical significance is limited.
Also, because the probability of a projected field galaxy increases with 
group area, one expects that that most discordant groups would be of low 
surface brightness. However, as several investigators have emphasized 
(Sulentic 1993, Arp 1995), this is not what is actually seen. A possible
explanation for this discrepancy is that Hickson's catalogue is not complete 
at low surface brightness (Hickson 1982, Prandoni \etal\ 1994). As a 
result, most low surface brightness groups (discordant or not) are not 
detected. Since previous analyses do not include this bias, the situation
needs to be reexamined.
 
In this paper we address two separate questions: 1) is the number of 
discordant redshifts in compact groups consistent with projection
effects, and 2) is there evidence for central concentration of
discordant redshifts? We improve upon previous work by making use
of a large homogenous catalogue of groups selected by an automated
procedure (Iovino et al., 1996). This allows us to predict the fraction of
discordant redshifts in an objective manner and with high statistical 
accuracy. These results are then applied to the Hickson catalogue of 
groups, which is 99\% complete in redshift measurements. To improve
statistics, we study both discordant quintets {\it and} discordant quartets 
in the catalogue. Our analysis explicitly includes incompleteness
effects, in both surface brightness and magnitude.

Our results confirm that projection alone can account for the high
incidence of discordant redshifts in compact groups. Contrary to
previous studies, we show that there is no evidence for central
concentration of discordant galaxies.

\section[]{Are Discordant Galaxies Centrally Located?}

Let us first consider the question of whether discordant galaxies 
(ie. galaxies with velocity within $1000\,\rm km \, s^{-1} $ of the median 
galaxy velocity of the group) fall preferentially closer to the centre 
of the group than do accordant galaxies, as has been suggested by previous 
work (Hammer \& Nottale 1986, Mendes de Oliveira 1995). 
As previous studies have suggested that discordant galaxies are
preferentially internal to the group, we consider the relative
numbers of {\i internal} galaxies. An {\it internal} galaxy is one 
whose center is located inside the smallest circle which contains the 
centers of the {\i other} galaxies. For the 100 groups in Hickson's
(1982) catalogue, 44 galaxies are discordant and 391 galaxies are
accordant, by the above definition. 43\% of the discordant galaxies 
are internal and 54\% of the accordant galaxies are internal. 
Clearly there is no preference for discordant galaxies to be internal.

We can also ask if the number of internal discordant galaxies
is consistent with a random distribution of galaxies on the sky.
For a set of $n$ galaxies randomly placed on a plane surface, 
the probability that any particular one of them will be internal is 
(from Appendix A) 
\be
	P_n = {(n-1)(n-2)\over n^2}~.
	\label{eq:prob}
\ee 
Multiplying this probability by the number of groups with 
one discordant redshift gives the predicted number of internal 
discordant redshift groups.

Table 1 lists data for the HCG groups. The columns are (1) Number of
galaxies in the group (accordant plus discordant), (2) number of groups
having one discordant galaxy (3) probability from Eq.~(\ref{eq:prob}), 
(4) predicted number of groups having one internal discordant galaxy,
(5) observed number of groups having one internal discordant galaxy,
(6) chance probability of finding at least that many discordant-redshift galaxies 
(from the binomial distribution). 

\begin{table}
 \centering
  \caption{Statistics of Internal Discordant Redshifts}
  \begin{tabular}{@{}lrrrrr}
   \hline \\[-12pt]
   $n$ & $N_{groups}$ & $P_n$ & $N_{pred}$ & $N_{obs}$ & $Prob$\\
   \hline \\[-12pt]
   4 & 19 & 0.3750 & 7.1250 & 6 & 0.78  \\
   5 &  6 & 0.4800 & 2.8800 & 4 & 0.31  \\
   6 &  3 & 0.5556 & 1.6667 & 0 & 1.00  \\
   7 &  0 & 0.6122 & 0.0000 & 0 & ---  \\
   8 &  1 & 0.6563 & 0.6563 & 1 & 0.66  \\ 
   \hline
\end{tabular}
\end{table}

From the table it is clear that the observations are in
accord with the predictions of the random model. All differences are 
attributable to chance. From the consistent negative results of both 
these tests, we conclude that {\it there is no evidence for central 
concentration of the discordant galaxies.}

\section[]{The Frequency of Projections}

In order to analyze the frequency of discordant redshifts, we
make use of both the SCG and HCG catalogues. The SCG catalogue
employs selection criteria designed to match exactly 
those of the HCG
catalogue ({\it ie} richness, compactness and isolation, 
Hickson 1982) and is obtained applying these selection 
criteria to a database of $\sim 1,000,000$ galaxies up to 
mag in $B_j \sim 19.5$, obtained 
through COSMOS scans of $\sim 200$ UKST $b_J$ plates 
(MacGillivray and Stobie 1984). 
However, because the SCG's are found by a computer
algorithm, they are not affected by any subjective or visual bias.
That makes them ideal for estimating the probabilities of
chance alignments. On the other hand, very few redshifts are  
available for this sample, so the actual numbers of discordant
redshifts are not known. The HCG catalogue includes redshifts for
almost all member galaxies. The observed numbers of discordant
galaxies are therefore known, but biases such as the incompletness
of the catalogue at low surface brightness makes calculation of
the probability of chance alignments uncertain.

Our technique, therefore, is the following: The SCG catalogue 
is used to determine the probabilities of chance alignments, and
to study the factors which affect these probabilities.
The results are then applied to the HCG catalogue, to see
whether or not the observed frequencies of discordant redshifts
are compatible with the projection hypothesis. Obviously this
approach can only work if the probabilities calculated from the
SCG are applicable to the HCG catalogue. In this section we
discuss the method used to estimate the probabilities, and 
the factors affecting them. We find that the probability of
a group being a chance alignment depends sensitively on the 
surface brightness of the group, and that other factors are 
much less important. 

The probability $p$ that a triplet will form a discordant quartet
due to chance projection of a field galaxy was determined by
taking all the triplets in the SCG catalogue, without  
any surface brightness limit, and placing each
of them at 100 random positions in the sky ({\it ie} giving 
them random coordinates within the area of the galaxy catalogue). The SCG search 
algorithm was then applied to see how many times a quartet
was formed which satisfied the selection criteria. Since it is
very unlikely that a random field galaxy will have the same
redshift as the triplet, the ratio of the number of quartets
found in this manner to the number of triplets times the
number of random positions gives the probability.

Since the above method places the triplets at random locations
with equal prior probability, it does not take into account galaxy
clustering or large-scale structure. In order to examine the
effects of clustering, a second series of runs was performed
in which the results were weighted according to the average 
density of galaxies in the region where the triplet 
was placed. In this case, the prior
probability of a triplet being found in a region of galaxy
surface density $\rho$ was taken to be proportional to $\rho^3$
(because the probability of each galaxy in the triplet should
be proportional to $\rho$, neglecting clustering within the 
triplets). For each position, a weight
proportional to $\rho^3$ was computed. The final probability,
denoted $\tilde{p}$, is then the sum of the weights at successful 
locations (\ie\ where a quartet was made) divided by the sum of 
all the weights, and is an upper limit to the true probability, 
having considered independently the members of the triplets. 

Similar runs were made in which the program checked for the
formation of a quintet, by superposition of two field galaxies
on a triplet. Also, the SCG quartets were moved to random
positions to determine the probability of forming a quintet
by superposition of one field galaxy on a quartet.

\section[]{Analysis}

Using probabilities calculated from the SCG catalogue, we can now
predict the numbers of discordant-redshift quartets and quintets in
the HCG catalogue. Figure 1 shows the distribution 
with surface brightness of the SCG quartets and of the discordants 
quartets found by randomly positioning the SCG triplets.
\begin{figure}
\psfig{file=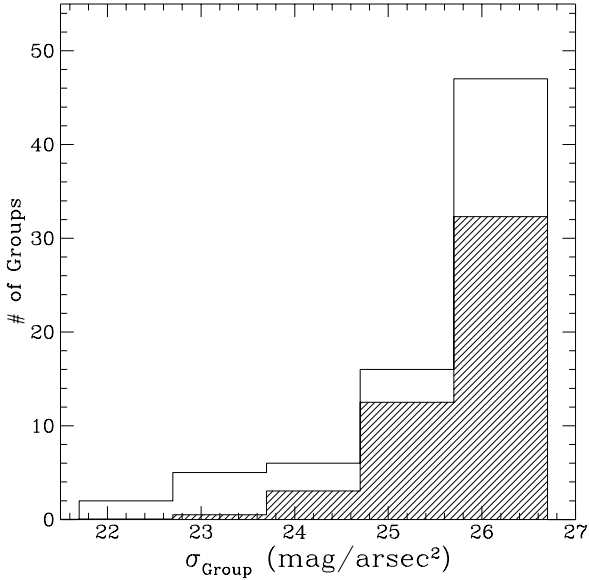,height=8.5truecm,width=8.5truecm}
\caption{ 
Histogram of the distribution as a function of surface brightness
of the number of SCG quartets (empty) and predicted discordant 
quartets (shaded). 
}
\end{figure}  
From the figure it is evident that the contamination rate 
increases as the surface brightness decreases.  
Because of this, the analysis is done separately in  
intervals of surface-brightness. 
This allow us to properly take into account the exact distribution 
in surface brightness of HCGs.    
To check for any possible dependence of the contamination 
rate on the magnitude limit of 
the groups considered, 
we have analyzed our data using two different magnitude limits. 
The results were found to be insensitive to the magnitude limit chosen. 

In order to minimize fluctuations due to
small number statistics, we use a new method of analysis: 
Consider the question of whether or
not the number of HCG quartets having one discordant member (\ie a
``3+1'' quartet) is consistent with chance projection. For every 
surface brightness interval, the following quantities are known:

\begin{tabbing}
~~~~\=$\hat{n}_3$~~~~~\= the number of triplets in the SCG catalogue \\
\> $\hat{n}_4$ \> the number of quartets in the SCG catalogue \\
\> $\hat{m}_4$ \> the number of quartets in the HCG catalogue \\
\> $\hat{m}_{31}$ \> the number of (3+1) discordant HCG quartets \\
\> $p$ \> the probability of projection onto a triplet \\
\end{tabbing}
where $p$ is the probability that a given triplet will form an
acceptable (according to the selection criteria) quartet due to random
projection on a single field galaxy. The quantity $p$ is computed
from Monte Carlo simulations, as described above, and is assumed to 
be accurately known.
The quantities with a hat are considered to be discrete random 
variables drawn from smoothly distributed parent populations.
The corresponding population means are denoted by the same symbol
without the hat and are unknown quantities. Other relevant unknown 
quantities are $q$, the probability of a given HCG quartet being
3+1 under the projection hypothesis, 
and $s$, the ratio of triplets/quartets
in the parent populations. We assume that this ratio is the same
for SCG and HCG populations (the selection effects will affect in the 
same way the detection of quartets and of triplets). 
For clarity we list below the unknown
variables and their allowable ranges:

\begin{tabbing}
~~~~\=$n_3$~~~~~\= mean no. of triplets in the SCG population~~\=$[0,\infty]$ \\
\> $n_4$ \> population mean of SCG quartets\>$[0,\infty]$\\
\> $m_4$ \> population mean of HCG quartets\>$[0,\infty]$\\
\> $q$ \> probability that an HCG quartet is 3+1\>$[0,1]$\\
\> $s$ \> population ratio of triplets/quartets\>$[0,\infty]$\\
\end{tabbing}

These unknown variables are not all independent, but, by virtue of
their definition, are connected by the following relationships:

\ba
	m_{31} & = & qm_4 \nonumber \\
	n_3 &  = & sn_4 \nonumber \\
	q   &  = & ps   \nonumber \\
	n_3 &  = & qn_4/p \label{rel}
\ea

They are related to the observables through the probability
distributions:

\ba
	P(\hat{n}_3|n_3) & = & {n_3^{\hat{n}_3}\over\hat{n}_3!} \exp(-n_3) 
		\nonumber \\
	P(\hat{n}_4|n_4) & = & {n_4^{\hat{n}_4}\over\hat{n}_4!} \exp(-n_4) 
		\nonumber \\
	P(\hat{m}_4|m_4) & = & {m_4^{\hat{m}_4}\over\hat{m}_4!} \exp(-m_4) 
		\nonumber \\
	P(\hat{m}_{31}|\hat{m}_4,q) & = & \Bigl(\begin{array}{c} \hat{m}_4 \\ 
		\hat{m}_{31} \end{array}\Bigr) q^{\hat{m}_{31}} 
		(1-q)^{\hat{m}_4-\hat{m}_{31}} \label{pdist}
\ea
Where the notation $P(a|b,c,\ldots)$ means the probabilty of obtaining 
$a$ given $b, c, \ldots$. 
The first three equations are Poisson distributions, which are appropriate
because the associated random variables have, in principle, no upper limit 
to their possible values. On the other hand, the distribution of
$\hat{m}_{31}$ is Binomial since that variable cannot exceed $\hat{m}_4$
(there cannot be more discordant quartets than there are quartets).

We wish to find the probability of observing $\hat{m}_{31}$ or more 3+1
quartets given the other known quantities, under the hypothesis of
chance projection. This is given by
\be 
	P = \sum_{x=\hat{m}_{31}}^{\hat{m}_4} P(x~|~q,\hat{m}_4)~ \label{extra1}
\ee
$x = q~\hat{m}_4$. Unfortunately, $q$ is not known, as one does not know the 
precise ratio $s = n_3/n_4$ of triplets to quartets. 
Therefore this uncertainty must be folded in the determination of 
P, giving: 

\be
	P = \sum_{x=\hat{m}_{31}}^{\hat{m}_4} P(x|\hat{n}_3,\hat{n}_4,
		\hat{m}_4,p)~. \label{prog}
\ee

An examination of the dependencies in Eq. (\ref{pdist}) shows that the
probability on the RHS of Eq. \ref{prog} can be factored:
\ba
	\lefteqn{P(x|\hat{n}_3,\hat{n}_4,\hat{m}_4,p)} \nonumber \\
	& = 
	\int P(x|\hat{m}_4,q)P(q|\hat{n}_3,\hat{n}_4,p) dq  \nonumber \\
	& = 
	\int \int P(x|\hat{m}_4,q)P(q|\hat{n}_3,n_4,p)P(n_4|\hat{n}_4) dq dn_4 
	\label{px}
\ea
where the integration is performed over the ranges of $q$ and $n_4$. Now
for given $n_4$ and $p$, $q$ is a function of $n_3$ by virtue of
Eq. \ref{rel}. Thus,
\be
	P(q|\hat{n}_3,n_4,p) = P(n_3|\hat{n}_3) {dn_3\over dq} = 
		P(n_3|\hat{n}_3){n_4\over p}~.	\label{pq}
\ee
This probability can, in turn, be transformed using Bayes' theorem:
\be
	P(n_3|\hat{n}_3) = P(\hat{n}_3|n_3) P(n_3)/P(\hat{n}_3)~.
\ee
where $P(n_3)$ is the prior probability of $n_3$, which is unity
because there is no prior information about this quantity, and
\be
	P(\hat{n}_3) = \int P(\hat{n}_3|n_3) dn_3 = 1~,
\ee
which follows from Eq. (\ref{pdist}). Thus we have
\be
	P(n_3|\hat{n}_3) = P(\hat{n}_3|n_3)~, \label{p3}
\ee
and similarly,
\be
	P(n_4|\hat{n}_4) = P(\hat{n}_4|n_4)~. \label{p4}
\ee
Substituting Eqs. \ref{pq}, \ref{p3} and \ref{p4} into Eq. \ref{px}
and then inserting the explicit probability distributions from
Eq. \ref{pdist}, we obtain
\ba
	\lefteqn{P(x|\hat{n}_3,\hat{n}_4,\hat{m}_4,p)} \nonumber \\ & = &
		\Bigl(\begin{array}{c} \hat{m}_4 \\ 
		x \end{array}\Bigr) {1\over \hat{n}_3!\hat{n}_4!}
		\int_0^1\int_0^\infty q^x (1-q)^{\hat{m}_4-x}
		\bigl({n_4q\over p}\bigr)^{\hat{n}_3} \nonumber \\
		& & \cdot ~\exp(-n_4 q/p) {n_4\over p} n_4^{\hat{n}_4} 
		\exp(-n_4) dn_4 dq \nonumber \\ & = &
	{\hat{m}_4!p^{-\hat{n}_3-1}\over x!(\hat{m}_4-x)!\hat{n}_3!\hat{n}_4!}
		\int_0^1 q^{\hat{n}_3+x} (1-q)^{\hat{m}_4-x} dq \nonumber \\
		& & \cdot \int_0^\infty dn_4 n_4^{\hat{n}_4+\hat{n}_3+1} 
		\exp(-n_4(1+q/p))~.
\ea
The $n_4$ integral is easily evaluated using the result
\be
	\int_0^\infty z^n \exp(-az) dz = {n!\over a^{n+1}}~,
\ee
which leaves,
\ba
	\lefteqn{P(x|\hat{n}_3,\hat{n}_4,\hat{m}_4,p)} \nonumber \\ & = &
		{\hat{m}_4! (\hat{n}_4+\hat{n}_3+1)! p^{-\hat{n}_3-1} \over
		x!(\hat{m}_4-x)!\hat{n}_3!\hat{n}_4!} \int_0^1 q^{\hat{n}_3+x}
		(1-q)^{\hat{m}_4-x} \nonumber \\
		& & \cdot ~(1+q/p)^{-\hat{n}_4-\hat{n}_3-2} dq~.
\ea
The remaining integral can be computed numerically, to give the desired
probability. We find, as expected that
\be
	\sum_{x=0}^{\hat{m}_4} P(x|\hat{n}_3,\hat{n}_4,\hat{m}_4,p) = 1~.
\ee
Probabilities for other cases (such as making 3+2 quintets by projection
of two field galaxies on a quintet) are determined in a similar manner.

The results are given in Table 2, in which the values of the relevant
observables are listed along with the probability $P$ that the observed
number of discordant systems, or more, would be found due to chance 
alignments with unrelated field galaxies.

\begin{table}
 \centering
  \caption{Probabilities of Discordant Redshifts}
  \begin{tabular}{@{}lrrrrrrrr}
   \hline \\[-12pt]
  \multicolumn{9}{c}{3+1 Quartets with $m < 15.5$} \\
   $\mu$ & $\hat{n}_3$ & $\hat{n}_4$ & $\hat{m}_4$ & $\hat{m}_{31}$ 
   & $p$ & $\tilde{p}$ & $P$ & $\tilde{P}$ \\
   \hline \\[-12pt]
   20.2 &   0 &  0 &  1 &  1 & 0.000 & 0.000 & 1.000 & 1.000 \\
   21.2 &   1 &  0 &  1 &  0 & 0.015 & 0.000 & 1.000 & 1.000 \\
   22.2 &   4 &  2 &  3 &  1 & 0.014 & 0.096 & 0.097 & 0.469 \\
   23.2 &  28 &  5 & 12 &  5 & 0.019 & 0.042 & 0.023 & 0.041 \\
   24.2 &  42 &  6 & 19 &  1 & 0.073 & 0.212 & 0.999 & 1.000 \\
   25.2 & 105 & 16 & 15 &  8 & 0.119 & 0.197 & 0.895 & 0.998 \\
   26.2 & 173 & 47 &  2 &  2 & 0.187 & 0.229 & 0.491 & 1.000 \\
   \hline \\[-12pt]
  \multicolumn{9}{c}{3+1 Quartets with $m < 14.5$} \\
   $\mu$ & $\hat{n}_3$ & $\hat{n}_4$ & $\hat{m}_4$ & $\hat{m}_{31}$ 
   & $p$ & $\tilde{p}$ & $P$ & $\tilde{P}$ \\
   \hline \\[-12pt]
   20.2 &   0 &  0 &  1 &  0 & 0.000 & 0.000 & 1.000 & 1.000 \\
   21.2 &   0 &  0 &  1 &  0 & 0.000 & 0.000 & 1.000 & 1.000 \\
   22.2 &   2 &  0 &  2 &  1 & 0.003 & 0.050 & 0.072 & 0.452 \\
   23.2 &   4 &  1 &  7 &  3 & 0.010 & 0.057 & 0.020 & 0.241 \\
   24.2 &   5 &  1 & 13 &  1 & 0.062 & 0.175 & 0.879 & 0.984 \\
   25.2 &  14 &  4 & 11 &  4 & 0.100 & 0.262 & 0.508 & 0.938 \\
   26.2 &  45 & 12 &  1 &  0 & 0.124 & 0.160 & 1.000 & 1.000 \\
   \hline \\[-12pt]
  \multicolumn{9}{c}{3+2 Quintets with $m < 15.5$} \\
   $\mu$ & $\hat{n}_3$ & $\hat{n}_5$ & $\hat{m}_5$ & $\hat{m}_{32}$ 
   & $p$ & $\tilde{p}$ & $P$ & $\tilde{P}$ \\
   \hline \\[-12pt]
   21.2 &   1 &  0 &  1 &  0 & 0.000 & 0.000 & 1.000 & 1.000 \\
   22.2 &   4 &  3 &  3 &  0 & 0.000 & 0.000 & 1.000 & 1.000 \\
   23.2 &  28 &  1 &  4 &  0 & 0.001 & 0.000 & 1.000 & 1.000 \\
   24.2 &  42 &  0 &  6 &  2 & 0.006 & 0.008 & 0.640 & 0.725 \\
   25.2 & 105 & 11 &  8 &  1 & 0.013 & 0.018 & 0.638 & 1.000 \\
   26.2 & 173 & 13 &  2 &  1 & 0.033 & 0.043 & 0.671 & 0.791 \\
   \hline \\[-12pt]
  \multicolumn{9}{c}{4+1 Quintets with $m < 15.5$} \\
   $\mu$ & $\hat{n}_4$ & $\hat{n}_5$ & $\hat{m}_5$ & $\hat{m}_{41}$ 
   & $p$ & $\tilde{p}$ & $P$ & $\tilde{P}$ \\
   \hline \\[-12pt]
   22.2$^a$ &   2 &  3 &  4 &  2 & 0.008 & 0.034 & 0.001 & 0.011 \\
   23.2 &   5 &  1 &  4 &  1 & 0.029 & 0.065 & 0.404 & 0.633 \\
   24.2 &   6 &  0 &  6 &  0 & 0.117 & 0.217 & 1.000 & 1.000 \\
   25.2 &  16 & 11 &  8 &  3 & 0.121 & 0.234 & 0.193 & 0.548 \\
   26.2 &  47 & 13 &  2 &  1 & 0.128 & 0.104 & 0.699 & 0.605 \\
   \hline
  \multicolumn{9}{l}{a: Seyfert's Sextet ($\mu = 21.52$) is included here}
 \end{tabular}
\end{table}

\section[]{Discussion}

From the tables it is clear that the numbers of discordant redshifts
found in the HCG catalogue are in accord with the projection
hypothesis in almost all cases. The exception is the highest surface 
brightness 4+1 quintet. Seyfert's sextet (HCG 79) has a surface brightness
of 21.52 mag arcsec$^{-2}$, which would place it in the $\mu = 21.2$
interval. However, there are no SCG quartets or quintets with
surface brightness in this range, probably due to the smaller area of the 
sky explored, so no statement can be made about 
probability for this interval (formally, the probability evaluates 
to 1.000). We have therefore conservatively included  
Seyfert's sextet in the 22.2
mag arcsec$^{-2}$ interval. The other group in this interval is Stephan's 
Quintet (HCG 92). With these two together, the probability of these
objects being due to uniform random projection is very small. If
clustering is taken into account, the probability increases to 1.7\%.
If Seyfert's Sextet is not included in the $\mu = 22.2$ interval, the 
probabilities $P$ and $\tilde{P}$ become 0.024 and 0.119 respectively.
Another well-known discordant group, VV 172 (HCG 55) falls in the
$\mu = 23.2$ interval. However, according to our analysis, the chance 
probability of finding a discordant group in this interval is high.

We conclude that practically all discordant redshifts in the
HCG catalogue can be explained by chance. On the other hand, it
would seem that Seyfert's Sextet, and to a lesser degree
Stephan's Quintet are unique objects, and that a resolution of their
nature must rest upon direct observations rather than statistical
arguments. Independent distance estimates, from the Tully-Fisher and 
$D_n-\sigma$ relations, have been obtained for galaxies in Stephan's 
Quintet (Kent 1981) and HCG 61 (Mendes de Oliveira 1995). In all cases 
the distances were found to be consistent with a cosmological 
interpretaton of the redshifts. The higher redshifts of galaxies in
Seyfert's Sextet make direct distance determinations difficult
(unless of course they are all much closer than their redshifts suggest).
It has been suggested (Hammer \& Nottale 1986, Mendes de Oliveira 1995)
that some of these groups may be cases of gravitational lensing,
but as we have seen above, there is no {\it statistical} evidence for 
this. Detailed observations of Seyfert's Sextet in particular might
indicate whether or not lensing plays a role in this group. 
We note for this group that the discordant galaxy is located 
close to the geometric center of the group, and is the smallest 
and faintest member. 

\section*{Acknowledgments}

PH thanks the Osservatorio Astronomico di Brera for its hospitality,
and the Canadian Natural Sciences and Engineering Research Council,
the Italian GNA, and NATO for financial 
support (NATO Grant No. CRG 920742). Both authors thank the referee, Gary 
Mamon, whose suggestions helped to improve the paper.

\appendix

\section[]{Probability of a galaxy being internal}

In this appendix we derive the probability $P_n$ that any galaxy
in a group of $n$ galaxies will be internal, under the assumption
that the galaxies are (uniformly) randomly distributed.
Each galaxy is represented by a point on a plane, corresponding to
the location of the galaxy's geometric centre.

Define the {\i boundary} of a set of $n$ points to be the smallest 
circle that contains all points. An {\i internal} point is a point 
which lies inside the boundary of the set formed by the other points. 
An {\i external} point is a point that is not internal.

It is easy to see that all points inside the boundary of a set are 
internal and all internal points lie inside the boundary: If a point 
is inside the boundary it can be removed without changing the boundary, 
so it is internal. If a point is internal it is inside the boundary 
of the remaining points, and so is inside the boundary of the full set.

It is evident that set (of 2 or more points) has either 2 or 3
external points. Two points suffice to define a circle by specifying a
diameter. Three points are sufficient in general. Since a circle has
zero width the probability of a 4th point falling on it is zero.

The probability of there being 2 external points in a set of $n$ points 
is 
\be
  P = 2/n. \label{eq:A1}
\ee
(Mamon, private communication) which can be deduced from Eqns 3, 4 
and 5 of Walke \& Mamon (1989).

We can now obtain the probability of a point being internal in a 
set of $n \geq 3$ points. If the set contains 2 external points, 
the probability that a given point will be internal is $(n-2)/n$. If the 
set contains 3 external points the probability is $(n-3)/n$. Weighting 
these two probabilities by the relative frequencies of 2 or 3 external 
points (from Eq. \ref{eq:A1}) gives
\ba
	P_n & = & {n-2\over n}P + {n-3\over n}(1-P) \nonumber \\
	    & = & {(n-1)(n-2)\over n^2}~.
\ea
\bsp

\end{document}